\documentstyle[preprint,12pt,aps,prabib]{revtex}
\tightenlines

\begin{document}

\draft

\preprint{$\ ^{\tt DFPD\ 97/TH/53\
(University\ of\ Padova)}$}

\title{ Spin observables for the $p d \leftrightarrow \pi^+ t$ reaction\\
around the $\Delta$ resonance}

\author{L. Canton, G. Cattapan, G. Pisent}
\address{Istituto Nazionale di Fisica Nucleare, Sezione di Padova,
         and Dipartimento di Fisica, Universit\`a di Padova,
         Via F. Marzolo 8, I-35131 Padova, Italy}
\author{W. Schadow}
\address{Department of Physics and Astronomy, Ohio University, Athens,
OH 45701, USA}
\author{J. P. Svenne}
\address{Physics Department, University of Manitoba, and
         Winnipeg Institute for Theoretical Physics, Winnipeg, MB,
         R3T 2N2 Canada}

\maketitle

\newpage

\begin{abstract}

The proton analyzing power $A_{y0}$ and the deuteron tensor 
analyzing power $T_{20}$ are evaluated for the $pd \leftrightarrow
\pi^+t$ process, in the energy region around and above the $\Delta$
resonance. These calculations extend a previous analysis of the 
excitation function and differential cross--section, based on
a model embodying one-- and two--body $p$--wave absorption mechanisms
and isobar excitation. The three--nucleon bound state and the $pd$
scattering state are evaluated through Faddeev techniques for both
the Bonn and Paris potentials. The spin variables exhibit a greater
sensitivity to the number of included three--nucleon partial waves
than the cross--sections, while the role played by the initial-- or
final--state interactions appears to be small. The results for the 
tensor analyzing power at backward angles show a non--negligible
dependence on the potentials employed, consistently with what
has been previously found for the cross--sections. 
The calculation of spin observables gives a clear indication that 
other reaction mechanisms (presumably $s$--wave two--body absorption) 
have to be included in the 
model, in order to reproduce the experimental data below the 
$\Delta$--resonance, in analogy with the simpler 
$pp\leftrightarrow \pi d$ process.

\end{abstract}


\newpage

\narrowtext

\section{Introduction}

      Pion absorption/emission processes on A=3 nuclei offer a unique
possibility for testing our present understanding of pion--baryon interactions
and reaction mechanisms with a minimum of {\it ad hoc} phenomenological
assumptions. It is now possible, in fact, to give a microscopic description of
both the nuclear bound state and the three--nucleon dynamics in the 
incoming (or exit) channel through modern few--body techniques.
\par
      A first step towards this goal has been accomplished in a recent
paper~\cite{cs97}, where pion emission/absorption reactions on tritium have
been studied in the energy region around the $\Delta$ resonance. The
elementary emission/absorption processes are given by the direct, 
one--nucleon mechanism, and by the two--body mechanism, proceeding
through $\Delta$ excitation. The former is described by a 
non--relativistic $\pi$NN vertex, the latter by a $\pi$N$\Delta$ vertex,
followed (or preceded) by isobar propagation, and a 
$\Delta$N$\leftrightarrow$NN transition mediated by $\pi$ and $\rho$
exchange. The three--nucleon bound and scattering states in the initial
and final channels have been evaluated through numerical solution of the
Alt-Grassberger--Sandhas (AGS)~\cite{ags} 
equations for realistic NN potentials,
by resorting to the Ernst-Shakin--Thaler (EST) expansion 
method~\cite{est,khp},
to produce a finite--rank representation of the NN interactions. 
With this representation, the $\pi$--production amplitude including 
the three--nucleon correlations can be
obtained by solving an integral equation where the 
off--shell extension of the plane--wave amplitude 
represents the driving term. This
approach is similar to a recent treatment of tritium 
photodisintegration~\cite{sasan}. 
The transition amplitudes have to be decomposed into 
three--nucleon partial waves, 
whereas the pion--tritium motion is 
represented by a three--dimensional plane--wave state.
The calculation has been performed in momentum space,
and details about the partial--wave 
formalism can be found elsewhere~\cite{cs97,casveca}.
\par
      The phenomenological parameters entering the above calculations
are essentially the coupling strengths and cut--offs at the meson--baryon 
vertices. These parameters had been tuned previously through the 
analysis of the simpler $\pi d \leftrightarrow pp$ 
reaction~\cite{can96,dor97}, over a large experimental database, including 
cross--sections and polarization observables. Hence, the analysis of 
Ref.~\cite{cs97} can be regarded as a parameter--free calculation, aiming at 
ascertaining to what extent the assumed model of the pion--nucleus dynamics 
can be extended from the two--nucleon to the three--nucleon system.
These calculations  have been compared with
experimental integral and differential cross--sections. 
The role played by the final-- or initial--state
interactions (FSI/ISI) has been studied, and the convergence with respect to 
the number of included partial--wave three--nucleon states has been tested. 
It turned out~\cite{cs97}  that the inclusion of higher partial waves 
and nuclear 
correlations have a comparable effect on the cross sections. Good 
results could be obtained in the considered energy region with both 
Bonn~\cite{bonnpot} and Paris~\cite{parispot} 
potentials, the latter producing a lower excitation function with respect to 
the Bonn {\sl B} potential, and slightly better results for the differential 
cross--sections at backward angles and high energies. Around the resonance, 
and for not too large angles, however, the results for the two potentials 
differ only in the normalization of the cross--sections, with the excitation 
function for the Paris potential being about 25$\%$ lower than the Bonn 
{\sl B} result.
\par
      In the present paper we extend the analysis to 
spin observables (such as the asymmetry $A_{y0}$, and the deuteron tensor 
analyzing power $T_{20}$), and to a larger energy region, in order to
get a deeper insight into the merits and limitations of the 
model employed for the $\pi$--three--nucleon dynamics.
\par
      For the $\pi d \leftrightarrow pp$ reaction, the polarization 
observables are known to be much more sensitive to the details of the 
dynamical model than the unpolarized cross 
sections\cite{can96,dor97,gmb}. 
For this reason, the large variety of spin  data has been 
never reproduced with complete success, 
in spite of the increasing complexity of the theoretical models. 
In particular, the proton analyzing power $A_{y0}$ is one of the most 
difficult observable to reproduce since it depends on 
both the magnitude and the relative phases of the 
intervening helicity amplitudes. A similar situation appears here 
for the $\vec p + d \rightarrow \pi^+ + t$ reaction. We investigated how
much this observable is affected by the partial--wave truncation,
and by the inclusion of nuclear correlations in the initial channel.
We found a more pronounced sensitivity to the number of included 
three--nucleon partial--wave states, with respect to the cross--sections. 
When the representation is enlarged to include 
18 two--body partial--wave states and 
S, P, and D orbital states in the intermediate $\Delta$ 
propagation, the theoretical results move towards the experimental points. 
The effects of the three--nucleon correlations in the initial state, on 
the other hand, turn out to be less important. As for $T_{20}$, one can 
obtain the general trend of the experimental data around the resonance, the
results exhibiting a moderate sensitivity to the chosen NN potential
at forward angles, and a greater sensitivity at large angles, 
consistently with the analysis of the unpolarized cross sections.
\par
      This model, in its present form, has been built 
for an analysis around the resonance region, where the intermediate 
$\Delta$N dynamics has to be treated explicitily. As well known, a theoretical
evaluation of the $\Delta$ width in the nuclear environment is still 
beyond present possibilities, since it would entail a consistent
solution of the coupled $\pi$NNN--$\Delta$NN--NNN problem. 
Lacking this theoretical
ingredient, we choose to parametrize the resonance width 
phenomenologically, starting from the experimental cross section
for the $\pi d \rightarrow pp$ reaction. This phenomenological
parameterization implicitly takes into account the non--perturbative aspects
of the intermediate $\Delta$ propagation within the two baryon subsystem; 
it proved successful both
in the description of the $\pi d \leftrightarrow pp$ data from 
threshold~\cite{cddo}
up to the resonance region~\cite{can96,dor97}, and in the three--nucleon 
calculations of Ref.~\cite{cs97}. Here, we extend the analysis of 
$pd \leftrightarrow \pi t$ processes beyond the resonance region for both 
the cross--sections and spin observables, and compare the outcome of 
this effective parameterization with the standard relativistic treatment 
of the $\Delta$ width, which applies to a free $\Delta$~\cite{ericweis,riim}. 
One finds that the former parameterization provides better 
differential cross--section in the resonance 
region, whereas the alternative description works better with increasing energy.
This suggests that non--perturbative 
effects in the off--shell $\Delta$ propagation are non--negligible 
in the resonance region, while their importance decreases at higher energies.
\par
 The data are not well reproduced in the low--energy region. This is 
particularly true for the tensor analyzing power at forward angles, 
where one fails even in reproducing the trend of the experimental points. 
This had to be expected, because in the present calculations only $p$--wave 
absorption mechanisms are considered. At low energies, one 
expects non--negligible contributions from $s$--wave pion absorption, involving
$\pi$ rescattering on a second nucleon. Such mechanisms play
an important role in the $\pi$--absorption process, 
and have to be taken into account in extending 
the present analysis from the high--energy to the low--energy region,
as has been shown in Ref.~\cite{cddo} for the simpler 
$pp\leftrightarrow \pi^+ d$ reaction.
\par
      The formalism relating cross--sections and polarization observables 
to the transition amplitudes as given in the present approach is 
reviewed in Section 2. The results of calculations are shown and discussed
in Section 3. Finally, Section 4 contains the summary and conclusions.

\section{Theory}

The amplitude for the $pd \leftrightarrow \pi^+ t$ process can be written
in terms of the matrix element

\begin{eqnarray}
A^{\rm TOT} = \,
{^{(-)}_{\cal \;\;\;S}} \langle{\bf q} \,, \psi_d|\, {\LARGE\cal A}\,
|\psi_{BS}\rangle_{\cal S}\, |{\bf P}^{\pi}_0\rangle \,.
\end{eqnarray}

\noindent
The states $|\psi_{BS}\rangle_{\cal S}$ and 
${^{(-)}_{\cal \;\;\;S}} \langle{\bf q} \,, \psi_d|$
describe the final three--nucleon bound state (BS) and the initial 
three--body scattering state, respectively, and are assumed to be properly
antisymmetrized. For both bound  state and scattering regimes,
we use herein the same three--nucleon states previously employed in
Ref.~\cite{cs97}.
Henceforth, we refer to that paper for any details about the
Faddeev--based calculation of the three--nucleon states.
The states $\langle{\bf q}|$ and $|{\bf P}^{\pi}_0\rangle$ are the
plane--wave states for the two fragments in the
asymptotic channels. The momenta $P^\pi_0$ and $q$ are the on--shell
momenta in the c.m. frame for the (outgoing) pion and
(incoming) nucleon, respectively.

In the operator ${\cal A}$ we specify the reaction mechanisms under
consideration. To avoid double countings, purely nucleonic intermediate
states must be avoided in ${\cal A}$, since the intermediate propagation
of three nucleons is taken into account to all orders when calculating
the three--nucleon dynamics in the final state.
These amplitudes are decomposed in three--nucleon partial waves, while
the pion--nucleus wave is kept three--dimensional.
We omit the details on the representation employed since
all this has been throughly discussed in previous
papers~\cite{cs97,casveca}.
Herein, we limit ourselves to mention that the three--body states
are defined in momentum space and the partial--wave decomposition
is discussed within the $jI$ coupling scheme.
The index $\alpha$ collectively denotes the whole set of quantum numbers,
namely spin, total angular momentum, and isospin of the pair ($s$, $j$,
and $t$, respectively) orbital, spin, total angular momentum, 
and isospin for the
spectator ($\lambda$, $\sigma$, $I$, and  $\tau$) and finally  the
three--nucleon total angular momentum, isospin and associated third
components, $JJ^z$ and  $TT^z$. To calculate spin observables we adopt
the helicity formalisms as introduced by Jacob and Wick~\cite{JW}.
The phase conventions, however, are those defined in Ref.~\cite{simonius}.
We transform the $jI$ coupling scheme of our amplitudes into a
channel--spin representation, where the angular momentum of the spectator
particle, $\lambda$, is coupled to channel spin $K$ (which is the
coupling of the spectator--nucleon spin, $\sigma_1$, to the deuteron
spin, $j$) to give the total angular momentum J. The transformation is

\begin{eqnarray}
|({ \lambda}  ({ \sigma_1}  { j}){ K}) J\rangle=
\sum_I (-)^{\sigma_1+\lambda+I}\hat K\hat I
\left\lbrace \matrix{j&\sigma_1&K\cr
\lambda& J&I}\right\rbrace
|({ j}  ({ \lambda} { \sigma_1}){ I}) J\rangle\,.
\end{eqnarray}

The amplitude in this new $\lambda$--$K$ representation
will be compactly indicated as

\begin{eqnarray}
A((\lambda, K)J,\mu_b)&\equiv \sqrt{4\pi\over 2J+1}
(-)^{{1\over 2}-J^z}
A(q,(\lambda K)J,-J^z,E)\,.
\end{eqnarray}

\noindent
We observe that in this representation the pion and nucleon
relative states of motion are treated differently,
and, for convenience, we prefer to discuss the amplitudes
in terms of the (inverse) pion absorption process. By doing so,
the initial state is {\it not} decomposed into partial waves,
while the final state is.
The quantization axis is parallel to
${P^\pi_0}$, thus it is the helicity axis for the incoming pion
(obviously, the helicity of the pion $\mu_a$ is zero).
For the target particle, the triton, we take the same axis,
but pointing in the opposite direction.
Hence, for the target particle, we have to rotate the frame of
$180^o$ around the normal to the scattering plane.
Because of this rotation the target helicity is $\mu_b=-J^z$,
and the phase factor appears in the equation above.
We construct the partial--wave helicity amplitude

\begin{eqnarray}
f^{/\ \mu_b}_{\mu_c \mu_d}(J)=
\sum_{\lambda, K}\langle\mu_c \mu_d J|\lambda K J\rangle
 A((\lambda K)J,\mu_b) \, ,
\end{eqnarray}

\noindent
where $\mu_c$ is the helicity of the outgoing nucleon,
and $\mu_d$ is the helicity of the deuteron.
Given that $\mu_a$ is zero, we use the symbol ``/'' in place.
The coefficients $\langle\mu_c\mu_d J|\lambda K J\rangle$ provide the 
Jacob--Wick transformation from the channel--spin to the helicity
representation with the
phase conventions adopted in Ref.~\cite{simonius}

\begin{eqnarray}
\lefteqn{\langle\mu_c\mu_d J|\lambda K J\rangle }  \nonumber\\
& = & (-)^{K-s_d+\mu_c} \,
C(s_c,s_d,K;\mu_c,-\mu_d,\mu_c-\mu_d) \,
C(K,J,\lambda;\mu_d-\mu_c,\mu_c-\mu_d,0)\, ,
\end{eqnarray}
where $C$ are the usual Clebsh--Gordan coefficients
in the notation of Ref.~\cite{casveca}.

The expansion in terms of reduced rotation matrices yields the angular
dependent helicity amplitudes

\begin{equation}
F^{{/\ }\mu_b}_{\mu_{c}\mu_{d}}(\theta) =
\sum_{J} { 2J+1\over 4\pi}f^{{/\ }\mu_b}_{\mu_{c}\mu_{d}}(J)
d^{J}_{(-\mu_b) (\mu_{c}-\mu_{d})} (\theta)\, ,
\end{equation}

\noindent
where $\theta$ is the scattering c.m. angle for the outgoing nucleon.
Of the twelve helicity amplitudes, symmetry principles
lead to 6 independent ones (we omit the angular dependence for brevity).
Thus

\begin{equation}
\renewcommand{\arraystretch}{1.2}
\begin{array}{lccl}
F^{{/\ } {1\over 2}}_{{1\over 2}\ {1}}& = & - \!\!&
F^{{/\ } -{1\over 2}}_{-{1\over 2}\ -{1}}\, ,\\*
F^{{/\ } {1\over 2}}_{{1\over 2}\ {0}}& = &   &
F^{{/\ } -{1\over 2}}_{-{1\over 2}\ {0}}\, ,\\*
F^{{/\ } {1\over 2}}_{{1\over 2}\ -{1}}& = & - \!\!& 
F^{{/\ } -{1\over 2}}_{-{1\over 2}\ {1}}\, ,\\*
F^{{/\ } {1\over 2}}_{-{1\over 2}\ {0}}& = & - \!\!& 
F^{{/\ } -{1\over 2}}_{{1\over 2}\ {0}}\, ,\\*
F^{{/\ } {1\over 2}}_{-{1\over 2}\ {1}}& = &   & 
F^{{/\ } -{1\over 2}}_{{1\over 2}\ -{1}}\, ,\\*
F^{{/\ } {1\over 2}}_{-{1\over 2}\ -{1}}& = &  &
F^{{/\ } -{1\over 2}}_{{1\over 2}\ {1}} \,.
\end{array}
\end{equation}

Finally, we obtain the helicity amplitudes for the $\pi$
production process from time--reversal symmetry.
Accordingly, the time--reversal amplitudes are related
by the equations

\begin{eqnarray}
F_{{/\ }\mu_b}^{\mu_{c}\mu_{d}}(\theta) =
(-)^{\mu_c-\mu_d+\mu_b}{|p_a|\over |p_c|}
F^{{/\ }\mu_b}_{\mu_{c}\mu_{d}}(\theta) \,,
\end{eqnarray}

\noindent
where on the left--hand side $\theta$ is the scattering angle for the
produced pion, while on the right--hand side it refers to the angle for the
outgoing nucleon. Also, $p_a$ is in our case $P^\pi_0$, while $p_c$ is $q$.
By means of these relations, we define

\begin{equation}
\renewcommand{\arraystretch}{1.2}
\begin{array}{lccl}
Z_1(\theta) &=& 
& F^{{/\ } {1\over 2}}_{{1\over 2}\ {1}} \,,\\*
Z_2(\theta) &=& 
-\!\!& F^{{/\ } {1\over 2}}_{{1\over 2}\ {0}}\,,\\*
Z_3(\theta) &=& 
& F^{{/\ } {1\over 2}}_{{1\over 2}\ -{1}},\\*
Z_4(\theta) &=& 
& F^{{/\ }{1\over 2}}_{-{1\over 2}\ {0}}\,,\\*
Z_5(\theta) &=& 
-\!\!& F^{{/\ }{1\over 2}}_{-{1\over 2}\ {1}} \,,\\*
Z_6(\theta) &=& 
-\!\!& F^{{/\ } {1\over 2}}_{-{1\over 2}\ -{1}} \,,
\end{array}
\end{equation}

\noindent
having factored out the momentum ratio since it is taken into account
at a later stage.
Well--known spherical tensor algebra~\cite{simonius} leads to the following
expression

\begin{eqnarray}
A_{y0}&=&-4 \, \frac{{\rm Im}(Z_1Z_5^*+Z_2Z_4^*+Z_3Z_6^*)}{I_0}
\end{eqnarray}

\noindent
for the analyzing power for the reaction
$\vec p + d \rightarrow \pi^+ + t$.
The quantity $I_0$ is defined as
\begin{eqnarray}
I_0=2 \, (|Z_1|^2+|Z_2|^2+|Z_3|^2+|Z_4|^2+|Z_5|^2+|Z_6|^2)
\end{eqnarray}

\noindent
and gives the differential cross section

\begin{eqnarray}
{d\sigma\over d\Omega}(\theta)&=& {c \over 2} \, I_0
\end{eqnarray}

\noindent
for the pion absorption reaction, or
\begin{eqnarray}
{d\sigma\over d\Omega}(\theta)
&=&
{c \over 6} \, \left( {P^0_\pi \over q} \right)^2 \, I_0
\end{eqnarray}

\noindent
for the inverse reaction.
The constant $c$ corresponds to the phase--space factor

\begin{eqnarray}
c={(2\pi)^4} {q\over P^0_\pi} {E_\pi E_{t} E_N E_d\over 
{(E^{tot})}^2} \,,
\end{eqnarray}

\noindent
with the relativistic energy of the fragments given by

\begin{equation}
\begin{array}{lcl}
E_\pi&=&\sqrt{m_\pi^2+{P^\pi_0}^2}\\*
E_{t}&=&\sqrt{M_{T}^2+{P^\pi_0}^2}\\*
E_N&=&\sqrt{M^2+{q}^2}\\*
E_d&=&\sqrt{M_D^2+{q}^2}\\*
E^{tot}&=&E_N+E_d=E_\pi+E_{t} \,.
\end{array}
\end{equation}

\noindent
Here, $M_D$, $M_T$, 
are the deuteron and tritium masses, respectively.

From the same helicity amplitudes, we have calculated also the deuteron
tensor analyzing power $T_{20}$

\begin{eqnarray}
T_{20}&=&\sqrt{2} \; \frac{(|Z_1|^2-2|Z_2|^2+|Z_3|^2-2|Z_4|^2+|Z_5|^2+|Z_6|^2)}
{I_0} \,.
\end{eqnarray}

\section{Results}

In a previous article~\cite{cs97}, it has been shown that
the energy dependence and angular distribution
of the $p d \leftrightarrow \pi^+ t$ reaction cross section
around
the $\Delta$ resonance could be reproduced
reasonably well with a meson--exchange isobar model
with the $\pi$--nucleon interaction mediated
by the $p$--wave $\pi$NN and $\pi$N$\Delta$ vertices.
The pion production/absorption diagrams included in that model
are shown in Fig.~\ref{figdiagram}.

In the same paper, the normalization of the cross section
was found to be quite sensitive to the NN
potential model employed 
(Paris, Bonn {\sl A}, or Bonn {\sl B}), while the angular
distributions were found to be less sensitive, with the exclusion
of the region at backward angles, where a non--negligible
dependence upon the nuclear potential has been shown.
At the resonance peak, the three--body dynamics in the
nucleon--deuteron channel (ISI) have been calculated
via a Faddeev--AGS computational scheme, and the effect
of the inclusion of a higher number of three--nucleon partial
waves has been also analyzed. In comparing the results, it turned
out that these two aspects were comparable in that they both
affected the unpolarized cross section by roughly the
same amount. Indeed, the Faddeev--AGS calculation of the
three--nucleon dynamics in the initial state gave a 4\%
effect in the cross section, and a comparable 4\% effect was
found in passing from a calculation including 82 three--nucleon
partial waves, to our largest calculation with 464 partial waves.

We refer to the article~\cite{cs97} for all the details about
the model, and for the list of partial waves
included in the various calculations. Herein, we limit ourself
to stress that the main difference between the 82-- and
464--state calculations is due to the inclusion in the latter
of $S$, $P$, and $D$ waves for the intermediate $\Delta N$ subsystem,
while in the former case only $S$--waves were retained.

In this section, we use the same amplitudes which have been
calculated previously in Ref.~\cite{cs97} and take the analysis
one step further by calculating the polarization observables
according to the methods described in the previous section.
The results for the proton analyzing power $A_{y0}$ are compared
in Fig.~\ref{figay0} with the data obtained in Ref.~\cite{cameron87}
for the isospin--related reaction $\vec p + d \leftrightarrow \pi^0
+ ^3$He at 350 MeV. Assuming charge--independence, the results
have to be equal.
The dashed line in the figure exhibits the result of the calculation
including 82 three--nucleon intermediate partial waves, and only
$S$ states for the intermediate $\Delta N$ system.
In this case, the results are very different from the
trend of the experimental data, however the dashed curve
is comparable in shape and magnitude with previous theoretical
results shown in Ref.~\cite{lolos82}.

The other two curves exhibit the results of the calculations
including 464 3N states and differ between each other by the
fact that the dotted line includes the effects of the
three--nucleon dynamics in the
initial state, while the full line does not.
In both cases the intermediate $\Delta N$ states
have been included up to $D$ waves.
In comparing the two curves, it is clear that
the consequences for $A_{y0}$ due to the
three--nucleon dynamics in the initial state are not very
large. Instead, the difference with the
dashed curve is a clear indication
that the polarization observables are
much more sensitive to the inclusion of a large, possibly
converged, set of states than what happens for the unpolarized
observables. In particular, an important aspect which cannot be
neglected is played by the $P$ and, to a lesser extent, by
the $D$ orbital states of the $\Delta N$ subsystem.
It was also found that orbital states higher than $D$ waves
give virtually no further change for this observable, at least
at the resonance energy.

In Fig.~\ref{figdiffbonn} the differential cross section
at forward and backward angles
is shown as a function of the parameter $\eta$, defined as
the pion--nucleus c.m. momentum, in units of pion masses
(times $c$). The experimental data have been obtained
at Saclay~\cite{mayer} for the $\pi^0$ production reaction.
The full circles represent cross section data at forward angles,
and the triangles are the corresponding data at backward angles.
The experimental results are compared with the theoretical
calculation assuming charge independence (implying a factor 2
between $\pi^+$ and $\pi^0$ production).
The full (dashed) line exhibits the calculation at forward
(backward) angles performed with the model introduced in
Ref.~\cite{cs97}, and correspond to results obtained under
the same circumstances as the results for $A_{y0}$ shown by
the full line in Fig.~\ref{figay0}.

These calculations have been tailored for the reproduction of the
pion production reaction in a region limited around the
$\Delta$ resonance. Indeed, the main limitation in the
energy range derives from the treatment of the width of the
$\Delta$ resonance, whose range of validity is restricted
in the region below $\simeq 2.5$ for $\eta$.
The explicit parametric form of the isobar width $\Gamma_\Delta$
is given in Ref.~\cite{can96}; it has been derived starting from
the shape of the experimental $\pi$--absorption cross section
on deuterons, $\sigma(E)$, by imposing the condition

\begin{equation}
\label{ritchie}
\sigma(E)={D\over (E-E_r)^2+{\Gamma_\Delta(E)^2\over 4}}\,,
\end{equation}

\noindent
plus the additional condition that
at the resonance peak $E_r$, the isobar width
coincides with the experimental value of 115 MeV
(this last condition fixes also the constant $D$).

The same energy dependence has been employed without further
changes in previous analyses for the
$\pi^+ d \leftrightarrow p p$ (Ref.~\cite{dor97}) and
$p d \rightarrow \pi t$ (Ref.~\cite{cs97}) reactions.
In this paper, as previously done in Ref. \cite{cs97},
we have used the same parameters (i.e., coupling constants
and cut--offs) defined in Ref.~\cite{dor97},
without further modifications.
However, in the analysis herein we extended the
energy range above the $\Delta$ resonance by going
up to $\eta$ $\simeq$ 4. In order to accomplish this,
we considered a different parameterization of the
isobar width,

\begin{equation}
\label{riim}
\Gamma_\Delta 
=
{2\over 3}
{f_{\pi N\Delta}^2
\over
4\pi}
{q^3\over m_\pi^2}
{M\over
\sqrt{s}}\,,
\end{equation}

\noindent
where $q$ is the pion momentum in the c.m. frame,
$\sqrt{s}$ is the invariant $\pi N$ mass,
$M$ and $m_\pi$ are the nucleon and pion masses, respectively,
and $f_{\pi N\Delta}$ is the coupling constant of the
$\pi N\Delta$ vertex.

This simple parameterization is commonly used in the analysis
of the $\pi N$ scattering processes~\cite{ericweis,riim}
and therefore refers to the width of a {\em free} $\Delta$.
It has the advantage of not being restricted to a limited
energy range, and hence can be employed in the region well
above the resonance peak where the previous
parameterization (Eq.~(\ref{ritchie})) breaks down.
A comparison between the energy dependence of
the two widths has been done in Ref.~\cite{can96}.

The dotted (dashed--dotted) line in Fig.~\ref{figdiffbonn}
show the forward (backward) cross section results calculated
with the isobar width expression given by Eq.~(\ref{riim}).
These results are spanning a wider energy region above the
isobar resonance, and they improve the description of the
reaction in the higher energy range, starting already from
$\eta$ $\ge$ 2.
In contrast, the results in the lower
energy region are best reproduced by the previous set of
calculations which employs the parameterization given by
Eq.~(\ref{ritchie}). At the present stage, there is no
definitive explanation for this behavior.
We may however argue that, as the pion momentum $\eta$
increases to a value around 2, the isobar in the
intermediate $\Delta$NN propagation tends to behave as a free--particle
 propagation, while at lower energy the importance of the
$\Delta N$ higher order interactions prevents a description in
terms of free isobar propagation.
If this is the case, the superior results obtained at lower energy
by the parameterization from Eq.~(\ref{ritchie}) can be understood
in the light of the fact that the $\Delta$ width obtained
directly from the experimental $\pi d\rightarrow pp$ excitation
function must include, albeit in some phenomenological and
approximate way, the $\Delta N$ higher order interactions,
while the baryon--baryon interactions are certainly excluded
when the width is parameterized from the $\pi N$ scattering data.

Fig.~\ref{figdiffparis} differs from the previous one in that
the Paris NN potential has been employed instead than the Bonn {\sl B}
one. Except for this change, for the four curves
the same symbolism of Fig.~\ref{figdiffbonn} has been adopted.
The forward peaked differential cross section are very
similar for the two potentials, the only
difference being in the normalization of the corresponding
curves. As already discussed in Ref.~\cite{cs97},
the results with the Paris potential are about 25\%  smaller than
the results obtained with the Bonn {\sl B} potential.
The results at backward angles show a greater sensitivity
to the NN interaction, since the differences in this case
cannot be attributed to a simple change in the normalization
of the results. At lower energy, the results at backward angles
obtained with the Paris potential reproduce better the
trend of the experimental data. At higher energy,
the calculations at backward angles  for both potentials
fail to reproduce the dip around $\eta \simeq 3.6$.
In a previous study~\cite{laget87}, the presence of this
``bump" in the backward--angle data has been explained in terms
of possible three--body pion production mechanisms.
Such mechanisms are not contemplated in the present analysis.

In Fig.~\ref{figt20forw}, the results for the deuteron tensor analyzing
power $T_{20}$ at forward angles are shown. The experimental results
are those reported in Ref.~\cite{mayer}.
The full and dotted lines represent the results obtained with
the Bonn potential, while the dashed and dashed--dotted
curves are similar calculations with the Paris potential.
Furthermore, the solid and dashed lines correspond to
the isobar width parameterized according to Eq.~(\ref{ritchie}),
while for the dotted and dashed--dotted lines Eq.~(\ref{riim})
has been used. Independently of the isobar width
and/or the nuclear potential employed the differences are not
very large.
The results at lower energy are very different from the experimental
data while at higher energy (for $\eta \ge 2$, where mainly
Eq.~(\ref{riim}) can be used) the reproduction is much better.
A possible explanation for this behavior can be attributed
to some low--energy production mechanisms still missing in 
the present analysis. A natural candidate is the mechanism
triggered by the $\pi N$ interaction in S wave. It is well
known that this mechanism is quite relevant in the $\pi d
\leftrightarrow pp$ reaction around the resonance~\cite{cs97,can96,dor97}
and of fundamental importance for the description of the same
reaction at threshold. It is possible, and indeed very likely,
that the same mechanism becomes more and more important at lower
energies also for the $pd\leftrightarrow\pi^+t$ process.
The discrepancies seen at lower energies
not only in Fig.~\ref{figt20forw}, but also in the forward--peaked
differential cross section 
(Figs.~\ref{figdiffbonn} and~\ref{figdiffparis}), 
could be possibly explained by such
mechanisms.

In Fig.~\ref{figt20back} the results obtained for $T_{20}$ at
backward angles are compared with the experimental data.
The symbolism of the figure is similar to that of the previous one.
It is evident that the backward angle results are much more
structured and difficult to reproduce than the forward results.
Here, the data are not reproduced well
in the entire energy region under consideration.
The theoretical results however show a certain structured shape
which is qualitatively similar to that evidenced by the
data. For this observable, similarly to what was previously observed for
the cross section, the backward--angle
results are quite sensitive to the NN potential employed.

\section{Summary and Conclusions}

In this paper we have carried further the analysis
of the $pd\leftrightarrow\pi^+t$ process initiated in Ref.~\cite{cs97}
by considering the spin observables.
In particular, we have studied the proton analyzing power $A_{y0}$
and deuteron tensor analyzing power $T_{20}$.

As is well known, $A_{y0}$ is very difficult to reproduce
because it depends not only on the magnitude of the various
helicity amplitudes, but also on the phases of these complex
quantities. We have found that
the three--nucleon dynamics in the initial channel has a modest
influence around the $\Delta$ resonance while in general the
convergence with respect to the partial--wave representation is
of great importance.
In particular, orbital $\Delta N$ states greater than $L=0$ 
definitely have to be
included. This effect on $A_{y0}$ is somewhat different
from the behavior of the unpolarized cross section,
where the three--nucleon dynamics and the higher partial waves
affect the results by roughly the same amount~\cite{cs97}.

We have considered the energy dependence of $T_{20}$ at forward
and backward angles, finding that $T_{20}$ at $0^o$ is not
very sensitive to the NN potential employed in the calculation,
while the results at backward angles are much more sensitive
to the nuclear potential,
except for the region at threshold. This behavior of $T_{20}$ is
similar to what has been observed previously for the differential
cross section.
Below the resonance, the $T_{20}$ results at forward angles
are in clear disagreement with the experimental results, while
above the resonance, the trend of the data is reproduced.
This behavior confirms the findings already obtained from the
analysis of the cross section at forward angles, and suggests
that there is some missing mechanism operating at low--energy.
From previous experience on the $\pi d\rightarrow pp$ reaction,
a very likely candidate is the process triggered by
the  $\pi N$ $s$--wave interaction.

Finally, we have compared the results obtained from two different
parameterizations of the $\Delta$ width. One has been obtained
from the phenomenology of the $\pi d\rightarrow pp$ cross section
while the other from a Chew--Low--type analysis of the $\pi N$ data.
While the second represents the standard parameterization of a single,
free isobar, the first includes phenomenologically the interaction effects
of the second nucleon on the isobar.
Around the $\Delta$ resonance the differential cross section is
better reproduced by the calculation with the former parametrization
of the $\Delta$ width, while the latter parameterization works
much better when the energy increases. This suggests that
at energies above the $\Delta$ resonance the intermediate isobar
propagates approximately as a free particle, while around the resonance
the $\Delta N$ interaction effects are not entirely negligible.

\begin{acknowledgments}

LC thanks B. Mayer for scientific communications.
GP thanks the Winnipeg Institute of Theoretical Physics
and University of Manitoba for support and hospitality in 
August--September, 1997. 
WS acknowledges financial supports from INFN and from the Deutsche
Forschungsgemeinschaft under Grant No. Sa 327/23-I.
JPS acknowledges financial support from a NATO Collaborative
Research Grant (CRG. 900551), with GP, at the very early 
stage of this work and the continuing financial
support from NSERC, Canada. JPS thanks also the University of Padova 
and INFN for hospitality in January--April, 1995 and July 1997.

\end{acknowledgments}

\newpage

\begin{figure}
\caption{\label{figdiagram}
Diagrams included in the present analysis. On
top, the $\Delta$--rescattering mechanism is
composed of the $\pi N\Delta$ vertex, the $\Delta$ intermediate
propagation, and the ${\Delta N}$ transition via $\pi$ and $\rho$
exchange.
On bottom, the direct $\pi NN$ mechanism is shown. For both mechanisms,
the three--nucleon correlations in the initial state are represented
by the oval on the left, while on the right the three--nucleon bound
state is denoted by the half oval.}
\end{figure}

\begin{figure}
\caption{\label{figay0}Analyzing power $A_{y0}$ of the reaction
$\vec p + d \rightarrow \pi^+ + t$ for incident
protons at 350 MeV . The dashed line includes only
intermediate $\Delta N$ S states, while the remaining
two curves include $S$, $P$, and $D$ intermediate $\Delta N$
states and 464 three--nucleon partial waves.
The dotted line contains the contribution of ISI,
while the full line does not. All the results
have been obtained with the Bonn {\sl B} potential.}
\end{figure}

\begin{figure}
\caption{\label{figdiffbonn}Differential cross section at $0^o$ and $180^o$
versus the parameter $\eta$ calculated with the Bonn {\sl B}
potential. The full (dashed) line represents
the forward (backward) cross section calculated with the
isobar width parameterized starting from the $\pi d$
absorption cross section,
while for the dotted (dotted--dashed) curves
the free isobar width has been employed.}
\end{figure}

\begin{figure}
\caption{\label{figdiffparis}Same results as in the previous figure
but with the Paris potential.}
\end{figure}

\begin{figure}
\caption{\label{figt20forw}Forward deuteron tensor analyzing power $T_{20}$
using Bonn {\sl B} (full and dotted lines) and Paris
(dashed and dotted--dashed curves) potentials.
For both the dotted and dotted--dashed lines
the free isobar width has been employed, while the
other two curves refer to the parameterization
derived from $\pi d$ absorption.}
\end{figure}

\begin{figure}
\caption{\label{figt20back}Same observable as in the previous figure but
for backward direction. The curves are defined with
the same symbolism.}
\end{figure}

\end{document}